\documentclass[%
 reprint,%
 amssymb, amsmath,%
 aip,pop,%
]{revtex4-1}

\usepackage{bm}%
\usepackage[colorlinks=true,linkcolor=blue]{hyperref}%
\usepackage{graphicx}
\usepackage{mathrsfs}
\expandafter\ifx\csname package@font\endcsname\relax\else
 \expandafter\expandafter
 \expandafter\usepackage
 \expandafter\expandafter
 \expandafter{\csname package@font\endcsname}%
\fi
\begin{document}

\title{A technique for plasma velocity-space cross-correlation}

\author{Sean Mattingly}%
\author{Fred Skiff}
\email{sean-mattingly@uiowa.edu}

\affiliation{212 Van Allen Hall, Department of Physics and Astronomy, University of Iowa, Iowa City, IA 52242}

\begin{abstract} 
An advance in experimental plasma diagnostics is presented and used to make the
first measurement of a plasma velocity - space cross - correlation matrix. The
velocity space correlation function can detect collective fluctuations of
plasmas through a localized measurement. An empirical decomposition, singular
value decomposition, is applied to this Hermitian matrix in order to
obtain the plasma fluctuation eigenmode structure on the ion distribution
function. A basic theory is introduced and compared to the modes obtained by
the experiment. A full characterization of these modes is left for future work,
but an outline of this endeavor is provided. Finally, the requirements for this
experimental technique in other plasma regimes are discussed.
\end{abstract}

\date{July 2017}%

\maketitle

\section{Introduction}

Kinetic modes, defined here as solutions to a plasma kinetic equation, have
recently been used to describe phenomena in astrophysical
plasmas\cite{Klein2012}, low density
plasmas\cite{Skiff2002,Souza-Machado1999,Diallo2005}, and fusion
plasmas\cite{Hatch2011,Terry2006,Terry2014}. Kinetic modes' importance is
partially due to the need to overcome the shortcomings of fluid and
magnetohydrodynamic descriptions that capture only a few modes in the full
plasma collective mode spectrum. This collective mode spectrum requires a
different mathematical approach that can give a general description of dynamics
in collisionless\cite{KAMPEN1955, CASE1959} and weakly collisional
plasmas\cite{NBS2004}. A problem remains that these kinetic modes, usually
being small and localized, are difficult to isolate experimentally. We seek to
introduce a new method that may be useful for detecting and measuring these
modes.

Often, kinetic modes can be best detected through phase space resolving
diagnostics. In our case, we use a phase space correlation measurement.
Correlations of plasma quantities have an extensive history, with recent
examples using density\cite{Tolias2015}, magnetic fields\cite{Bhat2015}, and
field - particle correlations\cite{Klein2016,Howes2017} to investigate
fluctuations, magnetic power spectra, and collisionless energy transfer in
turbulence respectively.  This illustrates a useful property of correlation
functions: depending on the plasma quantities correlated, they provide insight
into specific aspects of plasma behavior. 

The main velocity sensitive diagnostic used in this paper is laser induced
fluorescence (LIF). It fits the criteria of being phase space resolving and,
with our setup, is able to measure a phase space correlation of plasma ion
distribution fluctuations:
\begin{equation} C(\vec{x}_1, \vec{x}_2, \vec{v}_1, \vec{v}_2, \tau) = \langle
        \delta f(\vec{x}_1, \vec{v}_1, t) \delta f(\vec{x}_2, \vec{v}_2, t -
        \tau) \rangle_t \end{equation}
where $\langle \rangle_t$ denotes a time average and $\delta f = f - \langle f
\rangle_t$ is the phase space distribution function fluctuation. This
particular correlation can provide insight into plasma kinetic modes.  

This work stands on a foundation of previous LIF measurements of $C$.  They
employed a single laser, measuring $C$ for $v_1 = v_2$, to measure LIF at two
separate points along the laser beam\cite{Fasoli1994a} and to measure $C$ as a
function of $v_1 = v_2$ and with varying separation $x_1 -
x_2$\cite{Diallo2005}. Bicoherence spectra were also derived from these
measurements\cite{Uzun-Kaymak2006}. We build on these measurements by
introducing a localized measurement technique that measures $C$ for $x_1 = x_2$
as a function of two separate and adjustable velocities $v_1$ and $v_2$. By
varying these velocities, a matrix of cross correlations in velocity space may
be obtained.  

In this Paper, we introduce a technique combining the velocity sensitive
diagnostic LIF with a velocity space correlation. Thus we demonstrate with this
technique the first measurement of a plasma velocity – space cross –
correlation matrix. It is possible that the measurement, due to measurement
complications, is valid at some frequencies but not at others. We show how to
verify the measurement through symmetry properties of the matrix.  By applying
a singular value decomposition to this Hermitian correlation matrix, we obtain
the velocity space degrees of freedom of the plasma fluctuations as a function
of frequency.  We also compare the eigenmodes given by the singular value
decomposition to a basic kinetic theory.

Our goal here is to introduce a new measurement technique, independent of
velocity resolving diagnostic, that provides a new velocity space perspective
of plasma fluctuations. This technique can detect kinetic modes in plasma
through a localized measurement. We want to emphasize that while we demonstrate
this measurement with laser induced fluorescence (LIF), it is applicable where
any velocity sensitive measurement is available and a multipoint measurement
may be difficult. Examples of this include a satellite with numerous velocity
sensitive instruments, several collective Thomson schemes focused on one point
in a fusion plasma, trapped plasmas\cite{Anderegg1997}, or laser cooled
plasmas\cite{Strickler}. To further the goal of introducing a useful
measurement technique, we briefly discuss the criteria and pitfalls in applying
this measurement to other plasmas at the conclusion of this paper.  

The velocity sensitive diagnostic in this experiment, laser induced
fluorescence (LIF), is traditionally used to measure moments of the plasma
distribution function such as density, temperature, average velocity, and
higher. Moments are the easier measurements with this diagnostic.  The fact
that LIF is a velocity sensitive diagnostic means that fundamental
plasma properties can be measured with it. There is a history of intricate
measurements of fundamental plasma properties using laser induced fluorescence,
including early optical tagging measurements\cite{Stern1985}, wave – particle
interaction\cite{Fasoli1993}, phase space response to linear and nonlinear
waves and phase bunching\cite{McWilliams1986}, and plasma presheath
measurements\cite{Oksuz2001,Hood2016}. All of these LIF measurements surmount
the issue of photon statistics noise. In our experiment there is a twofold
contribution to noise. First, there is a large contribution ($> 90 \%$ of signal) from collision
induced background fluorescence. Second, LIF itself is limited to low photon
rates by low metastable state densities and thus we have issues with photon
counting statistics.

Correlations offer a way to remove this noise, in addition to providing a
measure of a useful plasma quantity. By using ensemble averaging, we can
isolate and eliminate the photon statistics noise. However, this is still a
technically difficult measurement to perform with LIF, which is why only the
above mentioned handful of correlation measurements with LIF exist. For this
reason, we quantify both the efficiency of background noise subtraction of the
collisionally induced fluorescence and whether we have counted enough photons
through symmetry properties of the velocity – space cross correlation matrix.
With this, we show how the measurement can be verified.

\section{Experimental Setup}

The experiment is performed on a cylindrical axially magnetized singly ionized
Argon (Ar II) plasma generated from an RF inductively coupled antenna of length
$2.3$ m and radius $\approx 2.5$ cm\cite{Diallo2005}. The magnetic field is .67
T. Langmuir probe measurements show ion and electron densities of $n \approx 9
* 10^9$ cm$^{-3}$ and $T_e \approx 9$ eV. LIF reveals $T_i \approx 0.08$ eV
though the distribution has significant deviations from a Maxwellian. Ion
neutral collisions have frequency $\nu \approx 500$ Hz. The plasma has ambient
acoustic and drift wave fluctuations above the thermal noise level, but they
are small due to convective stabilization. An antenna driven with white noise
drives fluctuations in the plasma. The plasma chamber contains two separate and
independently movable carriages with light collection optics for measuring LIF.
We leverage their independence to make this measurement.  

\begin{figure*} 
        \includegraphics{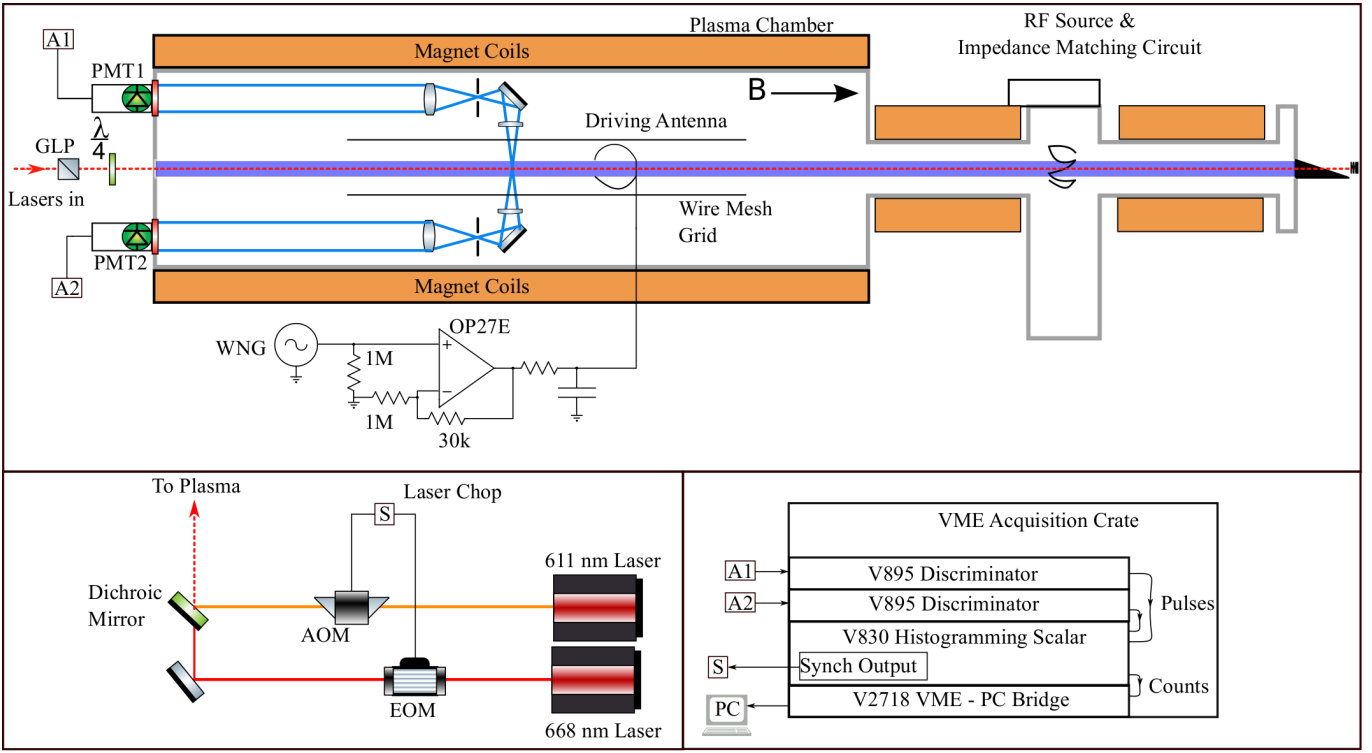} 
        \caption{Experimental
        setup. Top: The full plasma chamber setup. Two independently movable
carriages within the plasma chamber have collecting optics for gathering,
collimating, and sending out LIF light to PMTs 1 and 2. GLP is a Glan Laser
Prism and $\lambda / 4$ is a quarter wave plate. A single loop antenna is
upstream of the viewing volume and is driven with an amplified white noise
signal from the white noise generator (WNG). Bottom Left: Laser set up. The two
lasers are combined with a dichroic mirror with 650 nm cutoff (Thorlabs
DMSP650). Each laser is amplitude modulated according to a data acquisition
synchronized signal fed into an acousto-optical modulator (AOM) for the 611 nm
laser and an electro-optical modulator for the 668 nm laser. Bottom right: The
PMT pulse digitization and readout scheme. The signals from the PMTs enter
discriminators (V895), which digitize the PMT pulses. These pulses are
histogrammed with a V830, which also outputs a logic signal for laser
modulation according to the histogramming logic. Finally, the results are read
out to a PC via the V2718 and written to disk.} 
\label{FIG:SETUP} 
\end{figure*}

The measurement setup is shown in Fig.~\ref{FIG:SETUP}. Two separate lasers are combined
using a dichroic mirror, filtered for a single linear polarization with a Glan
laser prism, right hand circularly polarized with a quarter waveplate, and then
sent into the plasma. The exclusive right hand circular polarization eliminates
the left hand Zeeman splitting subgroup. Similarly, laser propagation parallel
to the magnetic field removes the $\pi$ perpendicular Zeeman
pattern\cite{Condon1959}.  

Each laser excites a separate independent laser induced fluorescence scheme.
Laser 1, a TOptica TA 100 Diode laser, excites the ArII metastable state
$^4$F$_{7/2}$ with 668nm to $^4$D$^\circ_{5/2}$ which decays to $^4$P$_{3/2}$,
emitting light at 442 nm; laser 2, a Matisse Rhodamine 6G Dye laser, excites
$^2$G$_{9/2}$ with 611nm to $^2$F$_{7/2}$ which decays to $^2$D$_{5/2}$, emitting
light at 461nm. Each laser absorption spectrum is broadened by the same ion
velocity distribution function, and each laser is velocity sensitive since the
laser bandwidth is $< 1$ MHz.

These two laser induced fluorescence schemes must be isolated - otherwise
transitions between the excited states can affect the cross correlation. There
are two major sources of transitions: collision induced transitions and atomic
transitions. Collision induced transitions may be immediately disregarded,
since their frequency at 500 Hz is very low compared to the atomic transition
frequency of 100 MHz. If an atomic transition pathway exists, however, the
cross correlation can be affected. We searched for a pathway experimentally by
modulating two separate lasers and looking for a beat frequency $f = f_1 \pm
f_2$. No beat frequency was found and so we believe that a pathway does not
exist for our particular laser induced fluorescence schemes we have chosen. Any
new pair of LIF schemes to repeat this measurement will need to be verified
similarly.  

Both sets of collection optics in the chamber are focused at the same point
with volume $0.20$ cm$^3$. Since the lasers are spatially combined but
independently tunable over velocity, and the optics are focused at the same
point, we are obtaining a measurement of the cross correlation function $C(v,
v', \tau)$ at two points separated in velocity space.

\section{Measuring the Velocity – Space Cross – Correlation}

We can now measure the velocity – space cross – correlation matrix. The
measured matrix is a three dimensional matrix of velocity $\cdot$ velocity $\cdot$ time.
The first two dimensions, in velocity, are selected by each laser, while the
third dimension (time) corresponds to the $\tau$ of the cross correlation
$C(v_1, v_2, \tau)$.  

\begin{figure} \includegraphics{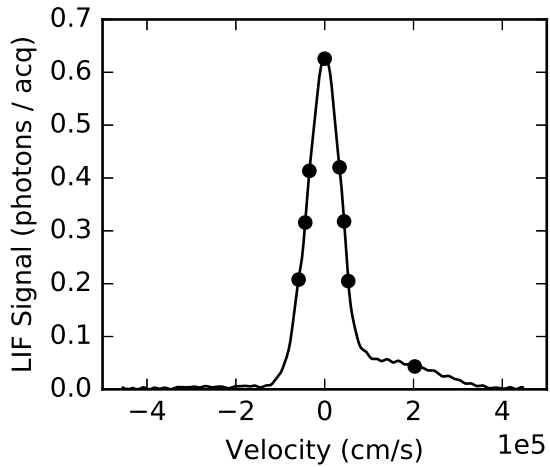} \caption{Deconvolved
        ion velocity distribution function measured by the two independent LIF
schemes.  The solid line is the ion velocity distribution function measured
with a traditional lock in amplifier setup and laser wavelength sweep. The anomalous
Zeeman effect has been removed via deconvolution. The circles are the
measurement points at the peak, 2/3 of the peak, 1/2, 1/3, and on the tail of the
distribution. This shows a typical IVDF in the plasma; the prominent tail at
high positive velocities is also visible.} \label{fig:IVDF}\end{figure}

We begin by measuring the full ion velocity distribution function with each
laser. We select points on the distribution corresponding to the peak, 2/3 of
the peak, 1/2 of the peak, 1/3 of the peak, and one point on the tail for seven
points total. Figure~\ref{fig:IVDF} shows the measured IVDF and the
measurement points on it. At each possible combination of the lasers, we
measure the time series data $f(v_1, t)$ from laser 1 and $f(v_2, t)$ from
laser 2. After demodulation with respect to 100 kHz laser chopping, the mean is
subtracted to provide the fluctuation $\delta f = f - < f >_t $. Cross
correlating and averaging with respect to $t$ gives the cross correlation
$C(v_1, v_2, \tau)$ for each of the selected velocities. This gives an 8 x 8 x
(2 N - 1) matrix where the first two dimensions are the velocities selected by
each laser and the last dimension is the time separation $\tau$.  

Accomplishing this measurement with LIF is technically difficult and requires a
large amount of time to overcome photon counting statistics noise, even after
the background noise from collision induced fluorescence is subtracted. The
data presented here took three weeks of continuous steady state plasma
operation and data acquisition. In addition to the reduction of noise through
this ensemble averaging, we suppress photon noise and background fluorescence
at several points: stray light filtering in the set up; background light
subtraction through LIF signal demodulation (this step removes background light
correlation as well); filtering via a Gaussian windowing function of the time
correlation; and the suppression of photon noise statistics through time
ensemble averaging. The last point is possible since the plasma is steady
state. Validating that this noise reduction works is important, so we
examine properties of the cross correlation matrix.  

Ideally, the cross correlation matrix $C(v_1, v_2, \tau)$ is symmetric such
that it equals $C(v_2, v_1, -\tau)$. However, the measured data are not
perfectly symmetric. There are errors in wavelength selection, and the magnetic
field induces slightly different Zeeman broadening in each laser's absorption
spectrum. This breaks the velocity space symmetry. Due to the low magnetic
field of 0.67 T and single circular polarization, the Zeeman subgroup spacing
is small compared to the measurement spacing. This is shown by the near
symmetry of the actual raw data matrix.

With the $C(v_1, v_2, \tau)$ matrix in hand, we can quantify the degree of
broken symmetry,  and thus, whether the noise suppression worked. The following
procedure gives where the measurement does have good symmetry as a function of
frequency. First, split the cross correlation matrix into symmetric and
antisymmetric components  $S, A = ½ C(v, v', \tau) \pm C(v', v, -\tau)$. Then
take the Fourier transform of $S$ and $A$ in order to obtain $\tilde{S}$ and
$\tilde{A}$, which are exactly Hermitian and antihermitian in the 2D velocity space by
construction. By a close form of the Fourier convolution theorem,
$\tilde{S}$ and $\tilde{A}$ are equal to the complex conjugate multiplied
symmetric and antisymmetric components of $\delta f (v, t)$ and $\delta f(v',
t)$, and the absolute value is the cross power spectrum. By comparing the cross
power spectra  of $\tilde{S}$ and $\tilde{A}$, we can verify 
the measurement is symmetric, and thus, our noise reduction techniques are
successful and external systematic effects are not too strong.  

\begin{figure} \includegraphics[scale=0.2]{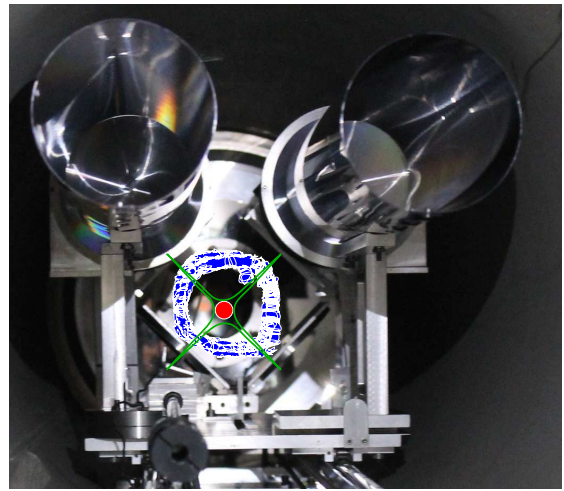}
        \caption{Photograph of the interior of the plasma chamber. This is from
                the perspective of the lasers that enter the chamber. The
                overlay describes important geometry of the experiment: red is
                the laser volume; green is the light collection volume; and
                blue is an idealized drift signature. The light collection
                volume is so shaped due to a pinhole.  This shows, physically,
                why we expect the $\pi/2$ phase contribution to the data matrix
        at $f^*$. Incidentally it also shows the two separate carriages inside
the plasma chamber. Collimated light going towards the PMT is going out of the
page.} \label{FIG:DRIFT_COMP} \end{figure}

\begin{figure}
        \includegraphics{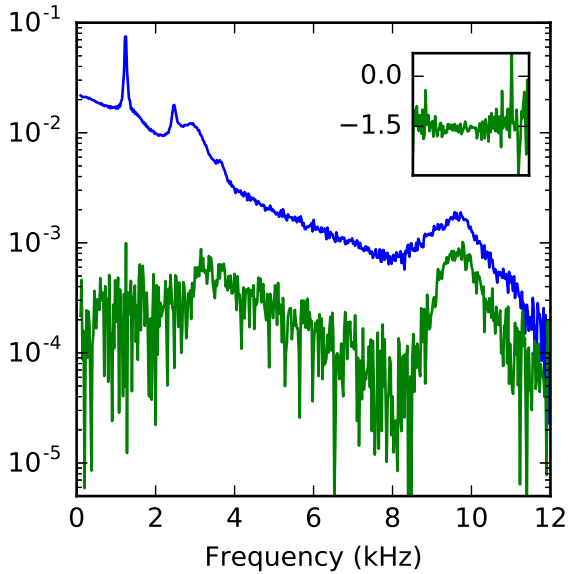}
        \caption{Representative power spectrum of a single velocity pair's time
                cross correlation.  The larger blue spectrum is the Hermitian
                component while the smaller green spectrum is the antihermitian
        component. The inset shows the phase of the antihermitian component
at the drift frequency and has the same abscissa as the outer figure. Reproduced
from S. Mattingly and F. Skiff, Phys. Plasmas 24 (2017), 10.1063/1.4996012 with 
permission from AIP Publishing.
}
        \label{FIG:HERM_ANTI}
\end{figure}

We choose a physically motivated example, drift waves, for verification of this
process.  Consider the physical set up in Fig.\ref{FIG:DRIFT_COMP}. The
periscopes are oriented at $90^\circ$ to each other and each has two light
collection volumes overlapping at the focus - where the lasers are located.
Pinholes ensure rejection of stray light outside the focus, but crucially
it is not complete rejection. While the lasers are at this localized point, the
optics collect light from the entire volume.  The drift wave amplitude peaks in
the gradient region of the plasma - higher up in the collection region and away from the
focus.  With this physical set up in mind, we quantify how the matrix symmetry
can be broken - and indeed identify at what frequencies the cross power
spectrum of the two lasers dominates, as opposed to other phenomena not in the
laser region.  According to the measured plasma parameters and adjusting for
the $k_\perp$ imposed by the wire mesh grid, we expect a drift frequency $f^*
\approx 10$ kHz. Additionally, we expect a phase near $f^*$ of $\pi / 2$, since
the strongest drift wave mode corresponds to the first Fourier mode
decomposition $e^{i m \varphi}$ where $m = 1$\cite{Horton1999} and the
periscope optics are oriented at $90^\circ$. In all, we expect the
antihermitian and Hermitian power spectra to approach each in other in strength
and the phase of the antihermitian spectrum to be $\pi / 2$ near $f^*$.
Figure~\ref{FIG:HERM_ANTI} confirms this. The antihermitian component
approaches the Hermitian component near $f^*$, and the inset shows the phase at
$\pi / 2$ verifies our expectations given the physical set up of the system.  

As a result, frequencies where the magnitude of the Hermitian component
dominate over the antihermitian component are where the measurement is
successful. This shows that not only have we suppressed background noise and
photon statistics, but that the Zeeman subgroup spacing is not too large. For
our particular measurement, this is true - to 10dB or better - for frequencies
below $f^*$.  Thus we have a measurement of the velocity - space cross -
correlation function and have quantified for which frequencies it is valid. We
have measured $C(v, v', \tau)$ at these frequencies. For these frequencies,
obtaining the velocity space degrees of freedom and their eigenmodes on the
plasma ion velocity distribution function is now possible.

\section{Velocity Space Degrees of Freedom}

We apply an empirical data transform, singular value decomposition
(SVD)\cite{Nardone1992}, in order to determine the velocity space degrees of
freedom of the plasma. We apply it to the Hermitian matrix $\tilde{S}(v, v',
\tau)$. Since this matrix is exactly Hermitian by construction, it is
diagonalizable. Thus the singular vectors returned by SVD are the eigenvectors
and the singular values are the absolute values of the eigenvalues. These
singular values are the relative strengths of their corresponding singular
vectors. Physically, the singular values can be interpreted as the strengths
of the different wave modes comprising the plasma fluctuations at a single
point in frequency. Write $\delta f$ as a sum of wave modes and expand $C (v,
v')$: 
\begin{align} &\delta f(v) = \sum_i^n a_i f_i(v)  \\ 
        C(v, v') &= \langle \delta f(v) \delta f(v') \rangle_t = \sum_i^n a_i
        f_i(v) \sum_j^n a_j f_j(v') \nonumber \\ 
        &= \sum_i^n |a_i|^2 f_i(v) f_i(v') + \sum_{i \neq j}^n a_i a_j f_i(v)
        f_j(v') \label{eqn:svd_modes} \end{align} 
Here, $f_i$ is a wavemode.  In the case of linearly independent plane modes,
the cross terms cancel out and we are left only with the simple basis $\sum_i^n
|a_i|^2 f_i(v) f_i(v')$.  

If the plasma modes are constituted of fluctuations that are linear and independent as
shown above, then singular value decomposition of the measured $C(v, v', f)$ matrix
would be very useful - it would diagonalize the matrix and pluck out
the modes $f_i(v)$ and their strengths $|a_i|^2$. These strengths $|a_i|^2$
correspond to the real valued singular values returned by SVD. They are the
velocity space degrees of freedom.  

However, the shape of fluctuations, especially in velocity space, is not necessarily
linear. In this case the above discussion is invalid since the cross terms of
Eqn.~\ref{eqn:svd_modes} would no longer cancel out. SVD's ansatz that the
basis is linear and orthogonal causes it to fail in this case. Determining and
applying a suitable transform for this case is a major avenue for future work
which we discuss in more detail in the final parts of this paper. For now, we
apply SVD to $C(v, v')$ keeping these drawbacks in mind.  

\begin{figure}
\includegraphics{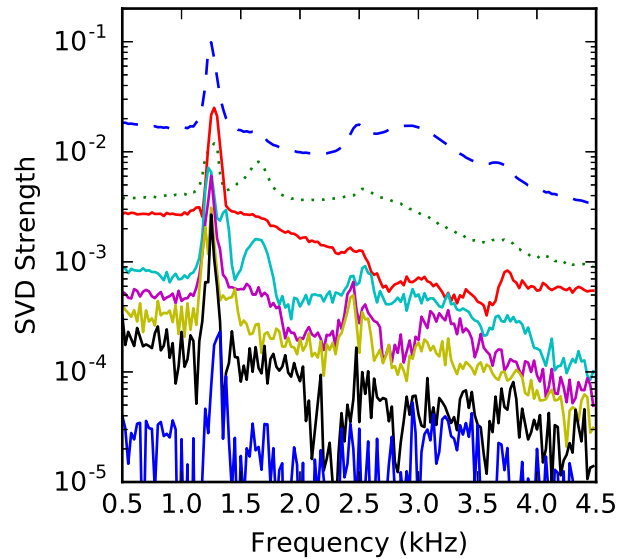}
\caption{Power spectra from connecting the singular values of the
                Hermitian matrix.  A separate SVD is run on each 2D velocity
                space and connecting the singular values across frequency gives
                this plot. The fact that these power spectra differ is evidence
                of distinct modes. The third strongest mode, in solid red,
                becomes second strongest around $1250\text{Hz}$ and then drops
                back to being third strongest. These are the velocity space
        degrees of freedom of plasma fluctuations as a function of frequency.
Reproduced from S. Mattingly and F. Skiff, Phys. Plasmas 24 (2017), 10.1063/1.4996012 with 
permission from AIP Publishing.
}
\label{fig:SVD_strengths}
\end{figure}

We apply SVD to the two dimensional velocity space matrix at every point in
the frequency spectrum.  Assuming continuity of the eigenvectors across
frequency, we connect the singular values across the spectrum. This process
gives spectra showing the relative strengths of the detected 
plasma fluctuation modes as a function of frequency and are shown in
Fig~\ref{fig:SVD_strengths}. These are the velocity space degrees of freedom.
We consider these spectra in greater detail to demonstrate spatial mode
separation from this localized measurement. Fully understanding these spectra
is a point of active and future work.  

The large peak at 1250 Hz matches the frequency of an ion acoustic bounded mode
in the plasma chamber with frequency given by $c_s / \lambda$, where $c_s$ is
the ion acoustic speed and $\lambda$ the wavelength of the bounded mode. Here,
this wavelength is $\lambda = 460$ cm, twice the length of the plasma column.
This peak has harmonics at 2500 Hz and 3750 Hz. The appearance of this mode in
all the SVD spectra shows a failure of the method at this particular wavelength
- the ion acoustic bounded mode is strong enough that it may be nonlinearly
scattering into the other modes. This causes the nonlinear terms of
Eqn.~\ref{eqn:svd_modes} to be nonzero. 

The peak at 1600 Hz and its harmonic at 3200 Hz are, we believe, due to a
longitudinal bounded mode in the wire mesh grid. The same calculation above for
the 1250 Hz peak, only with the dimensions of the wire mesh grid, gives these
frequencies. But this is due to a rough estimate; we do not actually know what
the boundary conditions are at the end of the mesh grid.  There are no
observations on a bounded mode like this to the authors' knowledge. Still, this
mode shows both advantages and limitations of the SVD method. First of all,
the peak at 1600 Hz is not apparent in the original power spectrum of
Fig.~\ref{FIG:HERM_ANTI}.  However, SVD clearly separates it into the second
and fourth spectra. This is an example of separating modes with this localized
measurement. This may also be a failure case - it may be a nonlinear mode that
SVD fails to separate.  Alternatively, it is possible that there are two modes
present which SVD is separating into these separate strengths. Finally, the
peak at 3200 Hz may not be a harmonic as the Lorentzian broadening (Q factor)
is not explained by calculations assuming reasonable reflection coefficients
from the open ended mesh grid. An intriguing possibility has been suggested
that it may be due to nonlinear interactions between the two 1600 Hz modes or
the 1600 Hz and 1250 Hz modes. Sorting all this out is a point of present and
future work.

The second strongest SV mode, colored with a green dotted line in
Fig.~\ref{fig:SVD_strengths}, becomes third strongest near 1250 Hz. We find
this by examining the singular value's corresponding eigenvectors as the
frequency is varied; specifically, we minimize the difference in eigenvectors
for adjacent points on the frequency axis. By enforcing this, we find that the
second and third SV modes switch in strength before and after the peak near
1250 Hz. This exemplifies SVD's lack of knowledge outside the frequency it is
applied.  Because it is a 2D transform applied repeatedly to 2D slices of a 3D
matrix, it only has information on the particular 2D space it is run on. Still,
we have pioneered a method of finding not only the relative strengths of these
modes from a localized measurement, but also how they change in relative
strength as a function of frequency. A point of future work is determining a
transform that can incorporate the third dimension of the matrix for
determining the mode strength as a function of frequency more reliably. This
will be especially useful when the mode crossing behavior is not sharply
defined, like it is in our case, but more gradual.

\section{Plasma Fluctuation Eigenmodes} 

Since the data matrix is Hermitian, singular value decomposition also gives
complex valued eigenvectors $f_i(v)$ corresponding to the singular values. In
the general non Hermitian case, these are the principal axes. We examine the
shape of these eigenvectors, and introduce a basic theory in an attempt to
explain some of them. The theory we introduce only meets with partial success; fully explaining
the modes is an area that we believe is rich in future work.


For comparison, we introduce a basic
kinetic theory based on electrostatic ion waves in a quasineutral plasma with
Boltzmann electrons.  We can obtain a useful form for $f_1$, the perturbed
distribution function, which can match up with some (but not all) of the modes
gives by the SVD analysis. First we write the Vlasov equation in Poisson
bracket form with a Bhatnagar - Gross - Krook (BGK)
operator\cite{Bhatnagar1954},
\begin{equation} 
        \partial f / \partial t + [f, H] = \nu (- f_1 + f_0 n_1 / n_0), 
        \label{eqn:vlasov} 
\end{equation}
and linearize, dropping second order terms.~$\nu$ is the collision frequency.
We use the ansatz $f_1 = Ae^{-i \omega t + i k_{||}z + i k_\perp(X +
\frac{v_\perp} {\Omega}sin\varphi)}$ with the gyrophase coordinate $X$ and
expand $f_1$ in Bessel functions of the first kind to get
\begin{equation} f_1 = \sum_{n=-\infty}^{\infty} f_n J_n(k_\perp v_\perp /
        \Omega). \end{equation}
where $\Omega$ is the ion cyclotron frequency and $v_\perp$ is the
perpendicular thermal velocity.  By solving the linearized Vlasov equation for
$f_n$, returning to $x$ from the gyrophase coordinate $x$, substituting the
$f_n$ obtained from Eqn.~\ref{eqn:vlasov} in the above $f_1$ expansion, and
integrating over $v_\perp$, we obtain 
\begin{eqnarray} f_1(v_{||}) = \sum_{n=-\infty}^{\infty} e^{-k_\perp^2 v_{T}^2
/ \Omega^2} I_n(\frac{k_{\perp}^2 v_T^2}{\Omega^2}) \frac{n_1}{n_0} k_B T_e
\nonumber \\ \times \frac{i k_{||}\frac{\partial f_{0}}{\partial p_{||}} +
(\frac{i n \Omega}{k_B T_\perp} + \frac{\nu} {k_B T_e})f_{0}}{\nu + i k_{||} v_{||} -
i \omega - i n \Omega}, \label{eqn:modes} \end{eqnarray} 
where $I_n$ is a modified Bessel function of the first kind. According to this
equation, at a given a frequency there are a range of modes present in the
plasma where each mode $f_1$ has its own pair $k_{||}$ and $k_\perp$. It is not
a dispersion relation.  Similarly, SVD resolves a subset of modes for each
given frequency and so we have separated the different spatial plasma modes
with a single localized measurement in velocity space.

\begin{figure*}
        \includegraphics{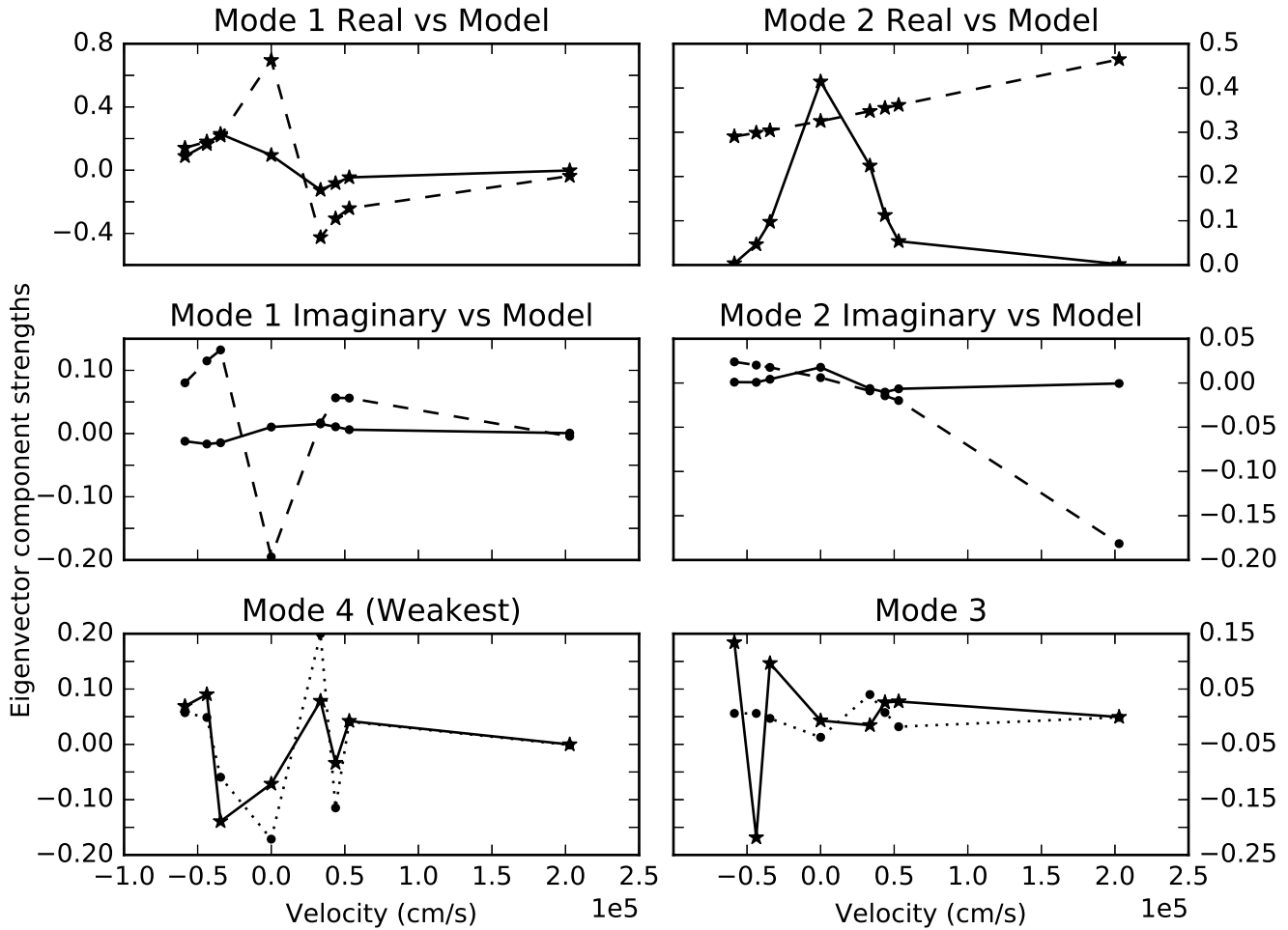}
        \caption{Eigenvectors of the four strongest singular value modes
                (in groups with strongest at top left; clockwise) at $f = 800$ Hz. In the first two
                modes, solid lines are derived from the experimental data while dashed lines are derived
                from the model of Eqn.~\ref{eqn:modes}. The third and fourth modes are derived solely from
                the experimental data; no fit is available. For them, the solid lines are the real component 
                while the dotted lines are the imaginary component. These results show that the strongest mode
                has the best fit, while the fit diverges for the tail point on the second strongest mode. 
                This shows where our theory works, and where it breaks down.
}
        \label{FIG:evecs}
\end{figure*}

We attempt to categorize the measured modes using Eqn.~\ref{eqn:modes}. By
evaluating Eqn.~\ref{eqn:modes} using the measured plasma parameters, the
collision frequency $\nu = 500$ Hz, and a large range of $k_\perp$ and
$k_{||}$; finding the norm of the difference between the obtained $f_1$ and the
data derived eigenvector; and finding the minimum of the resulting surface of
values, we obtain theory predicted values for the eigenvectors. It is
interesting to note that one may also fit for the collision frequency, though
we choose to keep it fixed at a realistic value. This could open another method
of collision frequency estimation given a well understood mode in the future.
The result of this process is shown for the two strongest eigenvectors in
Fig.~\ref{FIG:evecs} for $f = 800$ Hz. The data derived eigenvector for the
next two weaker modes are also shown, though we are not able to fit them. 

The strongest mode fits with the model for the tail and the sides of the
distribution, but not at the peak of the distribution. It resembles a typical
linearized ion acoustic wave response, only derived from the SVD analysis.  The fitted
parameters in Eqn.~\ref{eqn:modes} are $\lambda_{||} \approx 14.6$ cm and
$\lambda_\perp \approx 1.4$ cm. The second strongest mode does not fit as well
due to divergence from the tail of the distribution; the shown mode's fitted parameters are 
$\lambda_{||} \approx 460$cm and $\lambda_\perp \approx 15.3 $cm.

This process does not work for all the vectors returned by SVD. The weaker
modes are not explained at all by Eqn.~\ref{eqn:modes}. Understanding and
fitting these modes is a point of future work. Still, we have demonstrated
spatial mode separation from a localized velocity space measurement.

\section{Summary}

In this paper, we have presented a new measurement technique for measuring a
plasma fluctuation velocity-space cross-correlation matrix. We demonstrated the
technique on a laboratory plasma device and thus obtained the first measurement
of this matrix. By examining symmetry properties of the matrix, we verified the
measurement for a range of frequencies. We further verify it with comparison to
a prominent drift wave feature in our plasma. For these verified
frequencies, we applied a singular value decomposition to the Hermitian velocity
space correlation matrix.  We discussed several longitudinal bounded modes in the
plasma chamber and the wire mesh grid inside the plasma chamber, showing both
advantages and drawbacks of SVD. These drawbacks readily lead to future work.
In addition, we showed and discussed the eigenmodes on the plasma fluctuation
distribution function given by SVD as well. Finally, we introduced a basic
theory from a Vlasov equation with a BGK operator in order to make estimates of
expected eigenmode shapes.  This theory has apparent shortcomings, and only
works for some eigenmodes, but not for others.

\section{Future Work: Velocity Space Matrix Decomposition}

A full characterization of the modes given by this experiment is a major avenue
of future work. Finding and applying a theory based decomposition is
intertwined in this. With a suitable theory based transform the problems
inherent to SVD shown in the discussion of Eqn.~\ref{eqn:svd_modes} may be
solved. With this in hand, one could more accurately characterize the plasma
modes detected with the method presented here.

There are two major candidates, at the time of writing and to the authors' best
knowledge, for theory based transforms. The first is the Morrison $\mathscr{G}$
transform\cite{Morrison1994a}, which transforms the collisionless linearized
Vlasov - Poisson system onto a Case - van Kampen (CVK) mode basis. Applying a
transform such as this to the measurement presented in this paper requires a
great amount of experimental and theoretical work. First, the $C(v, v')$
matrix would need to be created for a much larger number than 8 measurement
points on the velocity distribution function. This is a technically difficult
endeavor for LIF mainly due to counting statistics. Second, applying the
$\mathscr{G}$ transform, a continuous transform, to the discrete data of this
experiment needs to be accomplished. A $\mathscr{G}$ based analogue to the
Nyquist sampling theorem and a generalized convolution theorem may be helpful
for applying the $\mathscr{G}$ transform to a discretely sampled set of data. 

Work on applying the $\mathscr{G}$ transform may also motivate the placement of
the measurement points on the ion velocity distribution function. Traditional
sampling theory dictates evenly spaced samples in order to perfectly
reconstruct an underlying sinusoidal signal. This may not be the case for
determining velocity space plasma modes from different measurement points on an ion
distribution function.  In the case of velocity space fluctuations the
underlying fluctuations have a different form, and so a measurement
``bunching'' on areas on the velocity distribution function may be necessary.
Study of the $\mathscr{G}$ transform may help determine where the measurement
points of $C(v, v')$ should be placed so that the $\mathscr{G}$ transform can
properly extract the underlying modes. 

Being able to reliably isolate CVK modes with this experimental technique and
the $\mathscr{G}$ transform would open experimental investigation of damping
mechanisms of CVK modes, which are not damped by Landau
damping\cite{Stix1992p308}. Another possibility, which requires a generalized
convolution theorem, is experimentally determining phase space fluctuation
spectra of density, electric field, and other plasma quantities using this
measurement technique and the $\mathscr{G}$ transform\cite{Morrison2008}.

The second major candidate for theory based transforms is through constructing
Hermite polynomials for a weakly collisional plasma\cite{NBS2004,
Souza-Machado1999}. This method uses Hermite polynomial expansions in the
Vlasov equation with a Lenard - Bernstein\cite{Lenard1958} collision operator
to find a discrete and infinite set of modes in velocity space. The Hermite
polynomial expansion is intrinsically discrete, which may make applications to
our dataset easier.  Still, similar to the $\mathscr{G}$ transform, work on the
Hermite expansion would be beneficial to identify the best measurement points
and apply it to the Hermitian data derived matrix presented in this paper.

The Hermite expansion technique may also provide a sort of bridge between the
current SVD methodology and a theory - driven transform. One may create a well
known mode, such as the ion acoustic longitudinal bounded mode, and generate
the Hermite polynomial adjoint matrix corresponding to the measurement points
on the distribution function. By left multiplying this adjoint matrix with the
Hermitian matrix at each frequency, then taking the singular value
decomposition, we can determine whether the modes isolated by SVD correspond to
the known modes. This is a point of future work.

Both of these transforms, $\mathscr{G}$ and Hermite, seek to sort out the
underlying eigenmodes of the system. In an infinite and homogeneous system, the
velocity-space cross-correlation eigenfunctions would be the plasma eigenmodes.
However, there is scattering between these infinite plasma modes due to long
correlation lengths, complicating the picture (and invalidating an analysis
like the SVD used earlier for them).  Projection onto either CVK modes or
kinetic modes via the $\mathscr{G}$ transform and Hermite polynomial
decompositions respectively could help to sort the situation out.

\section{Future Work: Application to other plasmas}

For this method to be applied to other plasmas, there are a few requirements
that the plasma system must fulfill. First, it must have two sets of
independently tunable, velocity sensitive diagnostics focused on the same
volume in the plasma. Ideally, they will be nonperturbative. We use LIF in this
experiment, but other ways are possible. For example, a satellite with multiple
Faraday cups tuned to separate velocities fits these criteria. Second, there
must be a way of taking an ensemble average to overcome counting statistics. We
accomplish this with a steady state plasma. But this is not the only way - a
good counterexample is a recent heterodyne method with LIF on a pulsed Hall
thruster\cite{Diallo2015}. Finally, the incoherent correlation length of the
time cross correlation of $f(v, t)$ and $f(v', t)$ must be short relative to a
single time series dataset but long relative to the sampling rate.

\section{Acknowledgements}

This work is supported by the US DOE under the NSF-DOE program with grant number
DE-FG02-99ER54543.

\end{document}